\pdfoutput=1
\documentclass[showpacs,superscriptaddress,floatfix,reprint]{revtex4-2}
\usepackage[latin1]{inputenc}
\usepackage[english]{babel}
\usepackage{graphicx}
\usepackage{amsmath}
\usepackage{amssymb}
\usepackage{amsfonts}
\usepackage{color}
\usepackage{bm}

\begin{document}
\title{Topological information device operating at the Landauer limit}
\author{A. Mert Bozkurt}
\affiliation{QuTech, Delft University of Technology, 2600 GA Delft, The Netherlands}
\affiliation{Kavli Institute of Nanoscience, Delft University of Technology, 2600 GA Delft, The Netherlands}
\affiliation{Faculty of Engineering and Natural Sciences, Sabanci University, Orhanli-Tuzla, Istanbul, Turkey}
\author{Alexander Brinkman}
\affiliation{MESA+ Institute for Nanotechnology, University of Twente, The Netherlands}
\author{\.Inan\c{c} Adagideli}
\email{adagideli@sabanciuniv.edu}
\affiliation{Faculty of Engineering and Natural Sciences, Sabanci University, Orhanli-Tuzla, Istanbul, Turkey}
\affiliation{MESA+ Institute for Nanotechnology, University of Twente, The Netherlands}
\affiliation{T\"UB\.ITAK Research Institute for Fundamental Sciences, 41470 Gebze, Turkey}
\date{\today}

\begin{abstract}
We propose and theoretically investigate a novel Maxwell's demon implementation based on the spin-momentum locking property of topological matter. We use nuclear spins as a memory resource which provides the advantage of scalability. We show that this topological information device can ideally operate at the Landauer limit; the heat dissipation required to erase one bit of information stored in the demon's memory approaches $k_B T\ln2$. Furthermore, we demonstrate that all available energy, $k_B T\ln2$ per one bit of information, can be extracted in the form of electrical work. Finally, we find that the current-voltage characteristics of topological information device satisfy the conditions of an ideal memristor.
\end{abstract}
\maketitle
Controlling dissipation in electronic devices is one of the major challenges in modern electronics. In fact one of the main aims of spintronics, and spin caloritronics in particular, is to study the interplay of heat, charge and spin transport in electronic devices~\cite{REF:Bauer2010}. This led to an impetus to develop spintronic devices, such as spin transistors and spin-dependent memory devices, as alternatives to their electron-charge based counterparts~\cite{REF:Datta1990, REF:Johnson1994, REF:Wolf2001, REF:Chappert2007, REF:Zutic2004, REF:Bauer2012, REF:Gaudenzi2018, REF:Puebla2020}, having low-energy dissipation and better storage capabilities. A parallel research was also conducted to better understand the thermodynamics of quantum coherent systems. Dubbed quantum thermodynamics, the field focused on how quintessential classical thermodynamics concept of heat engines, or the statistical mechanics concept of Maxwell's demon, would translate to the quantum domain. This research led to the study of quantum heat engines, where the working fluid is a quantum object or proposals of few-electron devices operating as "Maxwell's demons", which offers another alternative method of reducing heat dissipation and efficient energy storage (see Ref.~\cite{REF:Maruyama2009} for a review). Maxwell's demon implementations also became a central theme in the upcoming field of quantum information thermodynamics~\cite{REF:Deffner2019, REF:Auffeves2021}. 

Implementing a Maxwell's demon (MD) could also be attractive from the spintronics point of view, as it offers the possibility to control and even reverse dissipation in a device as long as another resource, namely demon's memory, is being consumed. Generating this resource in turn has a cost, stated by the Landauer's principle~\cite{REF:Landauer1961, REF:Bennett1982}: erasure of one bit of information requires a minimum amount of energy $k_B T\ln2$ to be dissipated, but this can be done at a later time.  
In order to have an observable and useful effect, one needs to design a MD to operate at a device scale, wherein macroscopic number of electrons need to be manipulated and perform the erasure as close as possible to the Landauer limit. However, most of the proposals \cite{REF:Anders2011, REF:Mandal2013, REF:Barato2013, REF:Strasberg2013, REF:Horowitz2013, REF:Deffner2013, REF:Park2013, REF:Rossnagel2014, REF:Strasberg2014, REF:Gemmer2015, REF:Pekola2016, REF:Kutvonen2016, REF:Lebedev2016, REF:Campisi2017, REF:Elouard2017, REF:Platero2017, REF:Ptaszynski2018, REF:Engelhardt2018, REF:Wright2018, REF:Croucher2021, REF:Croucher2022, REF:Lebedev2018, REF:Klaers2019, REF:Sanchez2019a, REF:Sanchez2019b, REF:Annby2020, REF:Josefsson2020}, as well as experimental demonstrations \cite{REF:Toyabe2010, REF:Berut2012, REF:Roldan2014, REF:Vidrighin2016, REF:Ciampini2017, REF:Camati2016, REF:Peterson2016, REF:Koski2014v1, REF:Koski2014v2, REF:Koski2015, REF:Chida2017, REF:Cottet2017, REF:Masuyama2018, REF:Naghiloo2018, REF:Erdman2018, REF:Kumar2018, REF:Ribezzi2019, REF:Paneru2020} are MD implementations comprised of a few particles at best. While a MD implementation that can operate at a device level was proposed by some of the us~\cite{REF:Bozkurt2018}, reaching the Landauer limit remains to be the main issue. 

In this Manuscript, we propose and theoretically investigate a novel spintronic device that operates as an information engine: a MD implementation which uses information entropy encoded in the spin degree of freedom to convert ambient heat into electrical work. The device is capable of operating on a macroscopic number of electrons {\em and} reaching the Landauer's limit, hence bypassing both issues of scalability and efficiency mentioned above. Compared to other Maxwell's demon implementations in other platforms, such as single-electron transistors~\cite{REF:Koski2015, REF:Dutta2017} and single qubits~\cite{REF:Pekola2016, REF:Elouard2020}, our proposed device has a natural way of scalability; by adjusting the device size, one can reduce or increase the number of nuclear spins, hence the energy output of the device. Furthermore, our proposed setup does not require any superconducting elements, which significantly increases the range of operating temperature of our device.
We show below that it is possible to extract all stored energy in the memory in the form of electrical work, completing the full thermodynamic cycle. This work extraction is achieved by connecting topological information device to an external load in order to generate electrical power. 

Furthermore, we demonstrate that topological information device can be utilized as a memristor. A memristor is an electronic circuit component \cite{REF:Chua1971} with a resistance that depends on the amount of charge that has flowed through it. By modeling the relation between the spin memory register and the electrical conductance of our topological information device we show that the current-voltage characteristics follow the ideal laws of memristance \cite{REF:Pershin2018} in a purely electronic device (i.e. without moving atoms). Therefore, the proposed topological information device is of fundamental interest as a model system for the interaction between information and energy, but could also lead to applications in energy efficient energy storage, in-memory computing as well as neuromorphic computing.  

\begin{figure}[tb]
\centerline{\includegraphics[width=\linewidth]{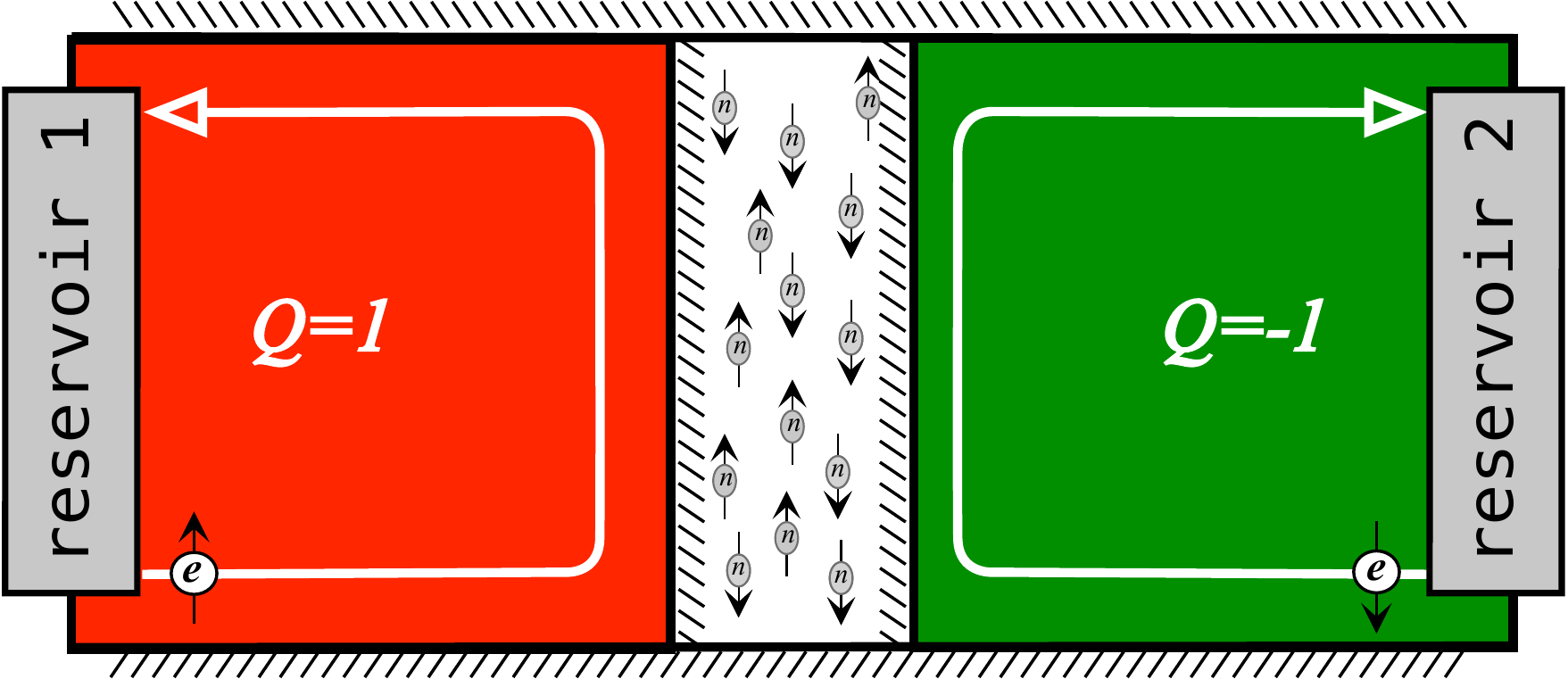}}\caption{Topological information device. The left lead is a quantum anomalous Hall insulator with spin-up electrons propagating in counter-clockwise direction, while the right lead is another quantum anomalous Hall insulator with spin-down electrons propagating in clockwise direction. The central region contains nuclear spins and/or magnetic impurities that allow electrons to transmit from one lead to the other lead via spin-flip process.}\label{FIG:qahi}
\end{figure}

\textit{Topological information device.---} Our proposed device is composed of two quantum anomalous Hall insulators with opposite spins and chirality with a junction area that contains nuclear spins and/or magnetic impurities as the MD memory (see Fig.~\ref{FIG:qahi}). A practical implementation of this device could for example be a magnetic topological insulator with a domain wall inside, separating the system into two distinct quantum anomalous Hall phases characterized by Chern numbers $\mathcal{Q}=\pm 1$. We discuss various alternative versions of its implementation in the Supplementary Material~\cite{FN:SuppMat}.

Without loss of generality, we assume that the left lead of our proposed topological information device has spin up electron edge states propagating counter-clockwise direction, while the right lead has spin down electron edge states propagating clockwise direction, as depicted in Fig.~\ref{FIG:qahi}. In the absence of any mechanism that can flip the spins of the electrons, the conductance vanishes because there are no available states for the electrons in the other lead. In fact, the only way for transmitting an electron through the device is to flip its spin via nuclear spins. In this way, any time a spin-flip occurs, a bit of information is erased and an electron is transmitted from one lead to the other. Consequently, the transmission of electrons results in Joule heating, in agreement with the Landauer's erasure principle. Conversely, polarized nuclear spins/magnetic impurities can be utilized to extract thermal energy from the reservoirs in the form of electrical power output. 

\textit{Model for topological information device.---} We now quantify our proposed device. The domain wall, where the nuclear spins are present, separates two regions with opposite Chern numbers $\mathcal{Q}=\pm 1$, therefore we have two co-propagating chiral edge modes at the boundary with opposite spins. We then project these co-propagating chiral edge modes at each side of the domain wall to the boundary~\cite{REF:Tokura2019} and obtain the effective Hamiltonian:
\begin{align}\label{EQN:Effective_Hamiltonian}
H = -i\hbar v_F \partial_s \sigma_0 + \lambda\sum_n \bm{\sigma}\cdot\bm{I}_n \delta(s - s_n), 
\end{align}
where $s$ denotes position along the junction that is perpendicular to the domain wall boundary, $v_F$ is the Fermi velocity of the chiral edge modes and $\sigma_0$ is the unit matrix in the electron spin subspace. The second term in Eq.~\eqref{EQN:Effective_Hamiltonian} describes the Fermi contact interaction between nuclear spins, described by the nuclear spin operator $\bm{I}_n$ for the $n^{\textrm{th}}$ nuclear spin at position $s_n$, and electrons with an interaction strength $\lambda = A_0 a^3$/S, where $A_0$ is the hyperfine interaction constant, $a$ is the lattice constant and $S$ is the cross section of the junction perpendicular to the domain wall. We stress that here, both spin species are co-propagating, as opposed to our earlier work where opposite spins are counter-propagating~\cite{REF:Bozkurt2018}. As a result of this feature, dissipation only takes place when there is an associated spin-flip.

We define mean polarization of the nuclear spins/magnetic impurities in the junction, $m = (N_{\uparrow} - N_{\downarrow})/2N$, which represents the state of the demon's memory. Here, $N_{\uparrow(\downarrow)}$ is the total number of up(down) nuclear spins and $N = N_{\uparrow} + N_{\downarrow}$ is the total number of nuclear spins in the junction. 

We find that, in the quasi-stationary state, the charge current flowing through topological information device is related to the mean polarization~\cite{FN:SuppMat}:
\begin{align}\label{EQN:current1}
I = eN\frac{dm}{dt}.
\end{align}
We see from Eq.~\eqref{EQN:current1} that the spin-flip mechanism is the only way current can flow through the system. In other words, there can be no electron transmitted through the system without an associated nuclear spin flip. In the presence of a constant applied voltage bias $V$, the rate of change of mean polarization is given by \cite{REF:Bozkurt2018, REF:Lunde2012, REF:DelMaestro2013}:
\begin{equation}\label{EQN:dmdt}
\frac{dm}{dt} = \gamma_0\frac{eV}{2\hbar} - \frac{m}{\tau(V,T)},
\end{equation}
where $\gamma_0 = \lambda^2/8\pi\hbar^2v_F^2$ is the effective interaction strength between an electron spin and a single nuclear spin, $\tau(V,T) = \left(\gamma_0 eV/\hbar\right)^{-1}\tanh(eV/2k_B T)$ is the nuclear spin dynamics time scale which depends on temperature $T$ and applied voltage bias $V$.

Eq.~\eqref{EQN:dmdt} shows that the memory composed of the nuclear spins/magnetic impurities and characterized by the mean polarization $m$ that can be erased by applying a voltage bias to the system via dynamical nuclear spin polarization. We find the mean nuclear spin polarization dynamics in the presence of DC voltage bias to be $m(t)=(m_0-\bar{m}){\rm e}^{-t/\tau}+\bar{m}$, where $m_0$ is the initial mean polarization and $\bar{m} = (1/2) \tanh(eV/2 k_B T)$ is defined to be the target mean polarization. 

We then use Eq.~\eqref{EQN:dmdt} and obtain the current-voltage characteristics of topological information device under DC voltage bias:
\begin{align}\label{EQN:dccurrent}
I = G_0 V\zeta\left[\frac{1}{2} - m\coth\left(\frac{eV}{2 k_B T}\right)\right],
\end{align}
where $G_0 = e^2/h$ is the quantum of conductance, $\zeta = 2\pi N\gamma_0$ is the effective interaction strength of topological information device. Note that the second term in Eq.~\eqref{EQN:dccurrent} reverses the dissipation: even in the absence of voltage bias, i.e. $V=0$, finite nuclear spin polarization and finite temperature generates finite charge current. This also holds in the absence of a temperature gradient within the system which can naturally generate a heat current. Therefore, we conclude that even if the electronic subsystem is completely at equilibrium, a charge current can be induced due to nuclear spin flips. On the other hand, the first term is the dissipative component of the charge current in accordance with the Landauer's erasure principle.

\textit{Memory erasure.---} The process of erasing the information stored in the memory of topological information device corresponds to polarizing the nuclear spins/magnetic impurities. We find that an erasure process using constant voltage bias requires heat dissipation above the Landauer limit, even though one can minimize the heat dissipation with a proper choice of voltage bias~\cite{FN:SuppMat}. However, minimizing the heat dissipation down to the Landauer limit is possible for an adiabatic erasure process, similar to a Carnot cycle. This translates into adiabatic change of the applied voltage bias such that the amount of power required is minimized. We accomplish this by having a voltage profile such that the charge current flowing through the device vanishes, as the power generated by the device is proportional to the charge current. As the current depends on the rate of change of nuclear spin polarization (see Eq.~\eqref{EQN:current1}), we find that a voltage bias profile that satisfies $\dot{m} = 0$ minimizes the heat dissipated by topological information device. Using Eq.~\eqref{EQN:dmdt}, we obtain the required voltage to be
\begin{align}\label{EQN:adiabatic_voltage}
V^*(m(t)) = \frac{2k_B T}{e}\tanh^{-1}(2m(t)).
\end{align}

We then use the voltage profile given in Eq.~\eqref{EQN:adiabatic_voltage} and obtain the amount of heat dissipated during the erasure process as
\begin{align}\label{EQN:adiabatic_heat}
Q &= eN\int_{0}^{1/2}dm\, V^*(m),\nonumber\\
&= Nk_B T\big( \frac{1}{2}\ln(1-4m^2) + 2m\tanh^{-1}(2m)
\big)\Big|_0^{1/2},\nonumber\\
&= Nk_B T\ln2,
\end{align}
where we see that the amount of heat dissipated in the process of adiabatic erasure of the information stored in the memory of topological information device reaches the Landauer limit.

\textit{Work extraction.---} Work can be extracted by utilizing a blank memory. A fully polarized nuclear spin subsystem, i.e. a blank memory, induces an imbalance between spin up and spin down electrons by transferring spin angular momentum from the nuclear spin subsystem to electron subsystem. As left and right leads have opposite spins and chirality, this imbalance generates a charge current, which then can be extracted by applying a reverse voltage bias or attaching topological information device to an external electrical load. In this way, Maxwell's demon harvests available thermal energy from the reservoirs and power the load. In other words, topological information device operates as an engine.

We consider an external load with conductance $G_L$ attached to topological information device. We make use of Eq.~\eqref{EQN:dccurrent} and the current conservation law and obtain \cite{FN:Footnote1}
\begin{align}
G_L V = -G_0 V\zeta\bigg[\frac{1}{2} -
m\coth\bigg(\frac{|eV|}{2 k_B T}\bigg)\bigg],
\end{align}
where $V$ is the induced voltage by topological information device. Solving for the induced voltage for a given load conductance, we obtain%
\begin{align}\label{EQN:bqievolt}
V = \frac{2k_B T}{e} \tanh^{-1}\left(\alpha m\right),
\end{align}
where $\alpha = \zeta/(G_L/G_0 + \zeta/2)$. We note that the induced voltage given in Eq.~\eqref{EQN:bqievolt} is of the same form as the applied voltage bias for adiabatic erasure in the charging phase, given in Eq.~\eqref{EQN:adiabatic_voltage}. The only difference is the multiplicative factor in front of $m$, where the former has the factor $\alpha$ and the latter has the factor $2$. In the limit of vanishing load conductance and hence vanishing power (reversible process), we have $\alpha \rightarrow 2$. In this case, we have the open circuit voltage $V_{\textrm{open}}$, which is directly proportional to the thermal energy $k_B T$, and solely relies on temperature. Conversely, in the device we previously introduced~\cite{REF:Bozkurt2018}, the open circuit voltage exhibits a linear dependency on the parameter $\zeta$. For currently available devices, we estimate $\zeta\sim 10^{-4}$ based on a device with approximately $N\sim 10^7$ nuclear spins and $\gamma_0 \sim 10^{-12}$. As a result, measuring the open circuit voltage in experimental settings for the topological information device becomes more feasible.

Using the voltage given in Eq.~\eqref{EQN:bqievolt}, we obtain the amount of work extracted from fully polarized nuclear spins as~\cite{FN:SuppMat}
\begin{align}\label{EQN:WorkExtraction}
W= Nk_BT \frac{2\zeta^{-1}}{\left(2-\alpha\right)}\frac{G_L}{G_0}\bigg(\ln\big(1-\frac{\alpha^2}{4}\big)
+\alpha\tanh^{-1}\big(\frac{\alpha}{2}\big) \bigg).
\end{align}

\begin{figure}[tb]
\includegraphics[width=\columnwidth]{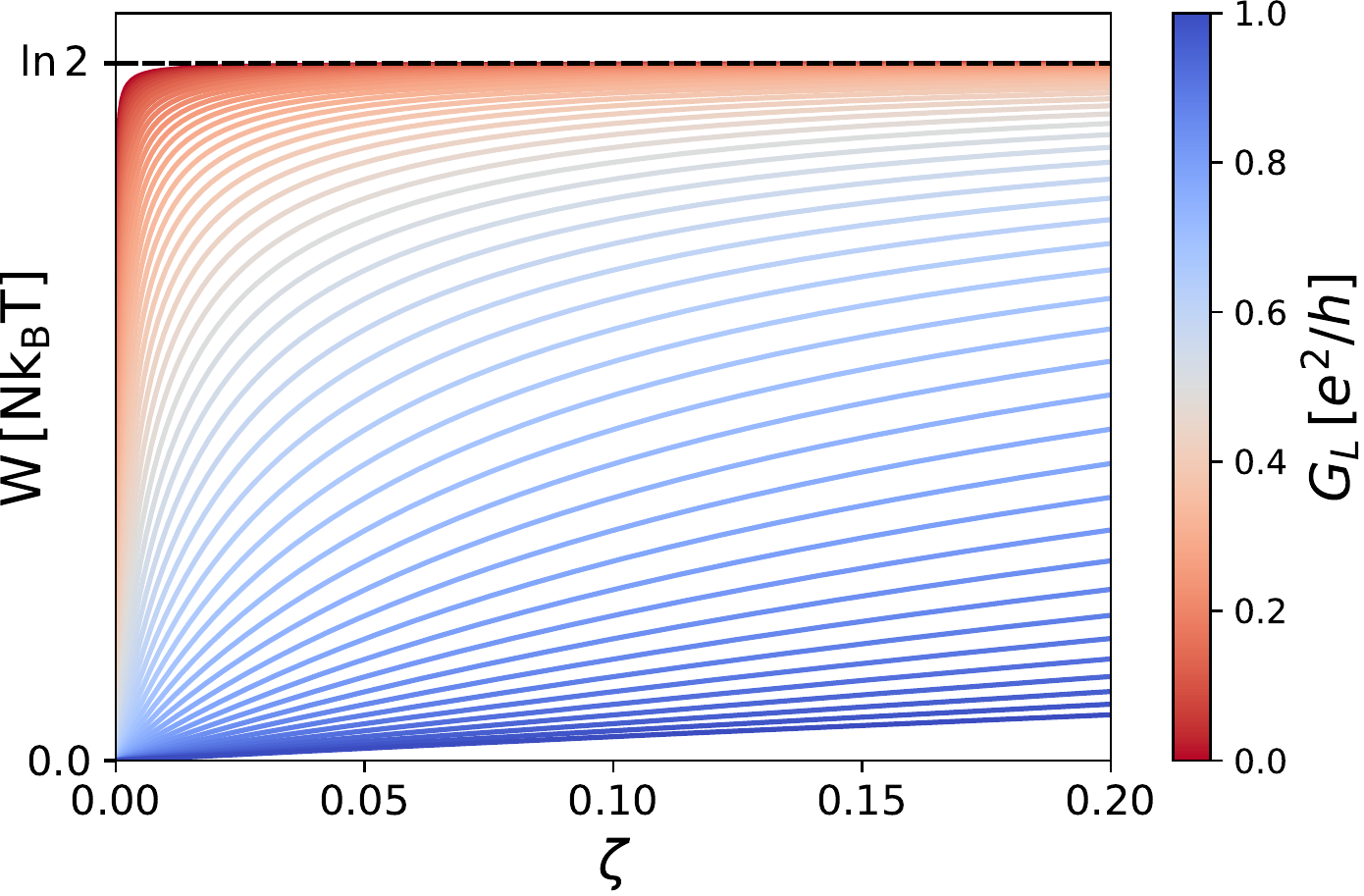}
\caption{The amount of work extracted per $N$ in units of $k_B T$ for topological information device vs. $\zeta$ with load conductance $G_L$ in units of conductance quantum $G_0$. The dashed line represents the Landauer limit on the extracted work per one bit of information.}\label{FIG:tid_multi_work}
\end{figure}

Fig.~\ref{FIG:tid_multi_work} presents the amount of work extracted per nuclear spin in units of thermal energy $k_BT$. We observe that in the limit of high load resistance, $G_L \ll G_0$, the amount of work extracted by topological information device reaches the Landauer limit more rapidly for a given $\zeta$ value. Furthermore, we examine the scaling behavior of the work extracted by topological information device with respect to $\zeta$ and find that the amount of extracted work per nuclear spin increases for increasing value of $\zeta$, before reaching the maximum value available, $k_BT\ln2$ per nuclear spin.

\textit{Memristor application.---} The first experimental realization of the memristor was reported in 2008 using a two-terminal device which is composed of a thin film of $\textrm{TiO}_2$ placed in between two platinum contacts~\cite{REF:Strukov2008}. Among multitudinous other platforms that were proposed and experimentally demonstrated \cite{REF:Wang2019}, spintronic systems are regarded as promising candidates for memristor applications \cite{REF:Pershin2008} as these are less prone to fatigue than systems with moving atoms. There have been various theoretical proposals \cite{REF:Wang2009, REF:Pershin2009, REF:Chen2010, REF:Grollier2016, REF:Dowling2022} and experimental demonstrations \cite{REF:Chanthbouala2011, REF:Chanthbouala2012, REF:Lequeux2016, REF:Mansueto2021} of memristive behavior in spintronic systems (see Refs.~\cite{REF:Grollier2020, REF:Shao2021} for an extended review). 

In general, any system with a memory that governs its resistance can be considered as a memristive system. These systems satisfy the following current-voltage relationship~\cite{REF:Chua1971}:
\begin{align}
I(t) = M^{-1}(x(t))V(t),
\end{align}
with $\frac{dx}{dt} = f(x,V)$, where $M$ is the memristance, $x$ is an internal state variable and $f$ is a function of the internal state variable $x$ and voltage $V$. On the other hand, as opposed to other memristive system, the resistance of an ideal memristor depends only on the amount of charge that has flowed through it and no other additional parameters \cite{REF:Pershin2018}. In other words, a memristor is defined as an ideal memristor, if the internal state variable $x$ corresponds to the charge $q$. We show that the current-voltage characteristics of topological information device meets this definition.

In our setup, the internal state variable $x$ corresponds to the polarization of the nuclear spins. Eq.~\eqref{EQN:current1} suggests that there is a one to one correspondence between the current flowing through the device and the rate of change of mean polarization in our system. This means that the change in polarization of the nuclear spins by an amount $dm$ provides means of tracking the amount of charge flowing through the device $dq$. We have
\begin{align}\label{EQN:charge_m}
dq = eN dm,
\end{align}
where we note that the direction of the polarity of the voltage applied defines the direction of the current. We see that the total amount of charge flowing through the device in a time interval $\Delta t = t-t_0$ is $q(t) - q_0 = eN\int_{t_0}^t dm = eN(m(t)-m_0)$ where $m(t_0) = m_0$ is the initial polarization and $q(t_0)=q_0$ is the initial charge. As the resistance of the system depends on the polarization of the memory, topological information device behaves as memristor in the limit of $|eV| \gg k_B T$. In this case, we rewrite Eq.~\eqref{EQN:dmdt} as
\begin{equation}\label{EQN:memristor_dynamics}
dm = \frac{\gamma_0}{\hbar} \bigg(\frac{eV}{2}
- |eV|m\bigg)dt.
\end{equation}

To find the memristance of the topological information device, we need to obtain its flux linkage-charge relationship $\varphi - q$. To that end, we rearrange Eq.~\eqref{EQN:memristor_dynamics} and integrate from a time $t^*$ at which the applied voltage $V(t)$ changes sign for the last time. In this case, we obtain the flux linkage $\varphi(t)$ as
\begin{align}\label{EQN:m_varphi}
\varphi(t) &= \frac{\hbar}{e\gamma_0}\int_{m(t^*)}^{m(t)} \frac{dm}{1/2 - \textrm{sgn}(V) m} + \varphi(t^*).
\end{align}

Without loss of generality, we consider a positive applied voltage bias. We then use the relation between charge and the polarization given by Eq.~\eqref{EQN:charge_m} and get the flux linkage as a function of charge as
\begin{align}\label{EQN:fluxcharge}
\varphi(t) &= -\frac{\hbar}{e\gamma_0}\log\left(1 - \frac{2}{eN}\frac{q(t)-q(t^*)}{1- 2m(t^*)}\right) + \varphi(t^*)\,
\end{align}
where $\varphi(t^*)$ depends on the history of the applied voltage bias from the initial time $t_0$ until time $t^*$:
\begin{align}
\varphi(t^*) = \sum_{n=1}^{n^*}    \frac{(-1)^n\hbar}{e\gamma_0}\log\left(1 +  \frac{(-1)^n}{eN}\frac{q(t_n)-q(t_{n-1})}{1/2+ (-1)^n m(t_{n-1})}\right)
\end{align}
where at each time $t_n$, the polarity of the applied voltage bias switches. Here, we define $t_{n^*} \equiv t^*$. We then use Eq.~\eqref{EQN:fluxcharge} to obtain the memristance $M(q)=d\varphi/dq$ of topological information device as
\begin{equation}\label{EQN:memristance_TID}
M(q) = \bigg[G_0 2\pi N\gamma_0 \bigg(\frac{1}{2} - \left( m(t^*) +
\frac{q-q(t^*)}{eN}\right)\bigg)\bigg]^{-1}.
\end{equation}

We emphasize that the flux linkage-charge ($\varphi - q$) relationship specified by Eq.~\eqref{EQN:fluxcharge} meets the three criteria for the ideal memristor: (i) Nonlinear; (ii) continuously differentiable; and (iii) strictly monotonically increasing \cite{REF:Georgiou2014}. Therefore, we conclude that topological information device functions as an ideal memristor in the limit of high bias voltage ($|eV| \gg k_BT)$.

\textit{Conclusion and outlook.---} In conclusion, we proposed and investigated a Maxwell's demon implementation in a spintronic setup. We demonstrated that topological information device is also an ideal memristor that can be used in electronic circuit applications. The main advantage of our device is the scalability of the MD memory (hence the energy storage capacity of the system) in conjunction with a minimum amount of heat dissipation during the memory erasure phase.

To provide an approximate estimation of the number of nuclear spins involved, we note that edge states in topological insulators can extend up to micrometers in length and span up to ten unit cells into the (insulating) bulk. Average number of nuclear spins per unit cell can reach up to 10, hence total number of nuclear spins that the edge states can interact can reach up to $N\sim 10^7$ nuclear spins.

We note that beyond the dominant nuclear spin relaxation mechanism, which is the interaction with electrons which was considered in this work, other relaxation mechanisms exist, such as dipolar interactions between nuclear spins or other correlations among nuclear spins~\cite{REF:Slichter1990}. We note that the timescale of such additional relaxation processes are estimated~\cite{REF:Khaetskii2003} to be orders of magnitude longer than the typical operation time of the topological information device. Correlations among nuclear spins are only important at sub-Kelvin temperatures~\cite{REF:Hsu2018}, significantly lower than typical device temperature.

We believe that our new device and the platform we proposed has the potential to advance the field of quantum thermodynamics by enabling researchers to investigate fluctuations more deeply. This can lead to a better understanding of concepts such as the Jarzynski equality~\cite{REF:Jarzynski1997} and Crooks fluctuation relations~\cite{REF:Crooks1999}, that were previously studied in electronic systems with single degree of freedom~\cite{REF:Saira2012}, hence we believe that our platform a good candidate as a testbed for theoretical ideas in the upcoming field of quantum thermodynamics.

\emph{Acknowledgments.---} We acknowledge helpful discussions with B. Pekerten, D. M. Couger, R. Salas and  S. K\"{o}lling. \.{I}.A. is a member of the Science Academy--Bilim Akademisi--Turkey; A.M.B. thanks the Science Academy--Bilim Akademisi--Turkey for the use of their facilities. This research was supported by a Lockheed Martin Corporation Research Grant.

\clearpage
\newpage
\onecolumngrid
\setcounter{equation}{0}
\setcounter{figure}{0}
\renewcommand{\theequation}{S\arabic{equation}}
\renewcommand{\thefigure}{S\arabic{figure}}
\renewcommand{\bibnumfmt}[1]{[#1]}
\renewcommand{\citenumfont}[1]{#1}

\vspace{0.5in}

\begin{center}
\begin{Large}
\textbf{Supplementary Material for \\
``Topological information device operating at the Landauer limit''}
\end{Large}\\[3pt]
A. Mert Bozkurt, Alexander Brinkman, and \.Inan\c{c} Adagideli
\end{center}

\section{Alternative Versions of Topological Information Device}

In this section, we introduce alternative versions of topological information device. The first alternative version is a hybrid system which contains two quantum anomalous Hall insulators with opposite spins and chirality, and a quantum spin Hall insulator that contain nuclear spins/magnetic impurities along a single edge. The helical edge states of quantum spin Hall insulator feature perfect spin-momentum locking. We make use of this property of the helical edge states in order to implement an alternative version of topological information device.
\begin{figure}[h]
\includegraphics[width=0.7\columnwidth]{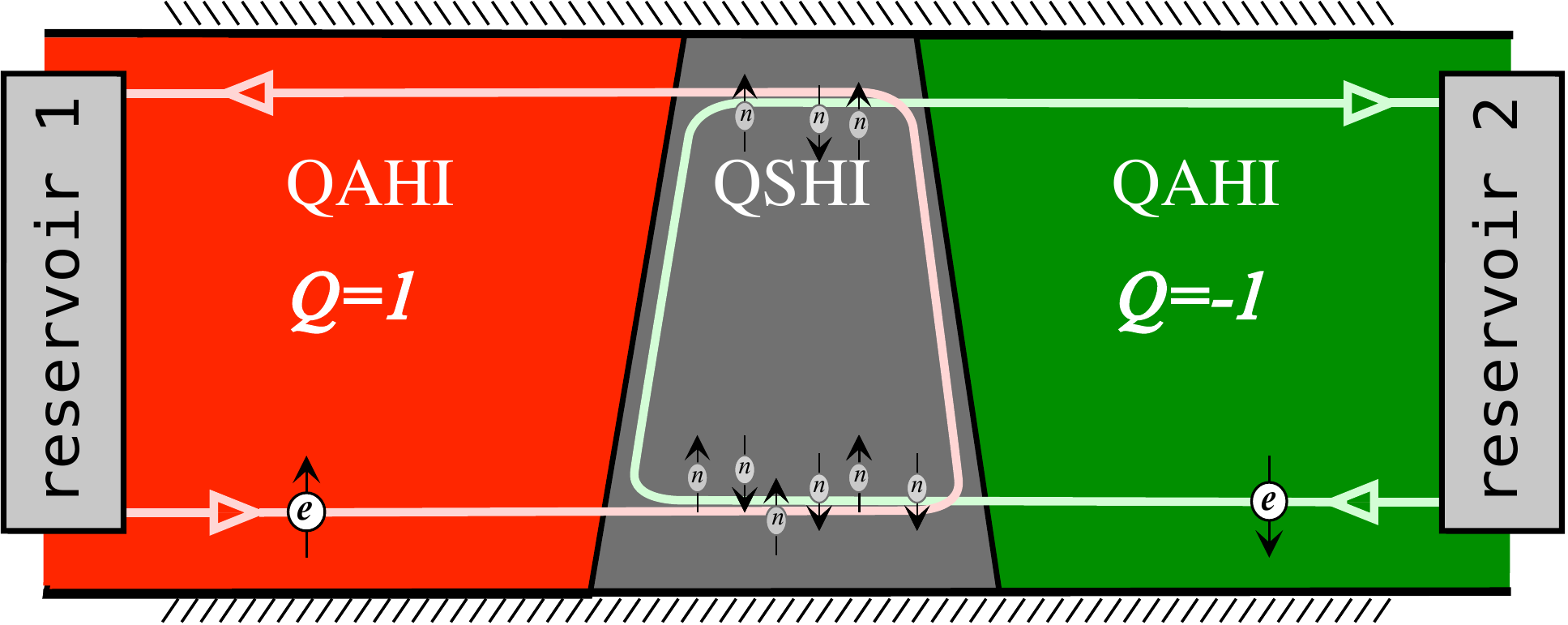}
\caption{First alternative version of topological information device. The leads are the same as in first version, but in this case, the central region is a quantum spin Hall insulator. In an ideal scenario for this alternative version, only the bottom edge contains nuclear spins/magnetic impurities.}\label{FIG:SuppMat:TID_v2}
\end{figure}

Similar to our proposal introduced in the main text, this version also allows transmission of electrons from one reservoir to the other only when a spin-flip occurs (see Fig.~\ref{FIG:SuppMat:TID_v2}). Due to perfect spin-momentum locking of the helical edge states, flipping the spin of the electrons leads to backscattering. As there are no available states for the spin-flipped electron in the lead that it is injected, the electron will be transmitted to the other lead. We note that transmission only takes place when an odd number of spin-flip occurs for a single electron. This is because even number of spin-flips leads to no change in electron spin (and also no change in total spin polarization of the nuclear spin), hence the electron backscatters into the same lead that it is originated, without any heat dissipation. 

The equations given in the main text for the original version of topological information device also applies to this alternative version. The only change then would be the total number of nuclear spins located at the bottom edge of the quantum spin Hall insulator.

\begin{figure}[t]
\includegraphics[width=0.68\columnwidth]{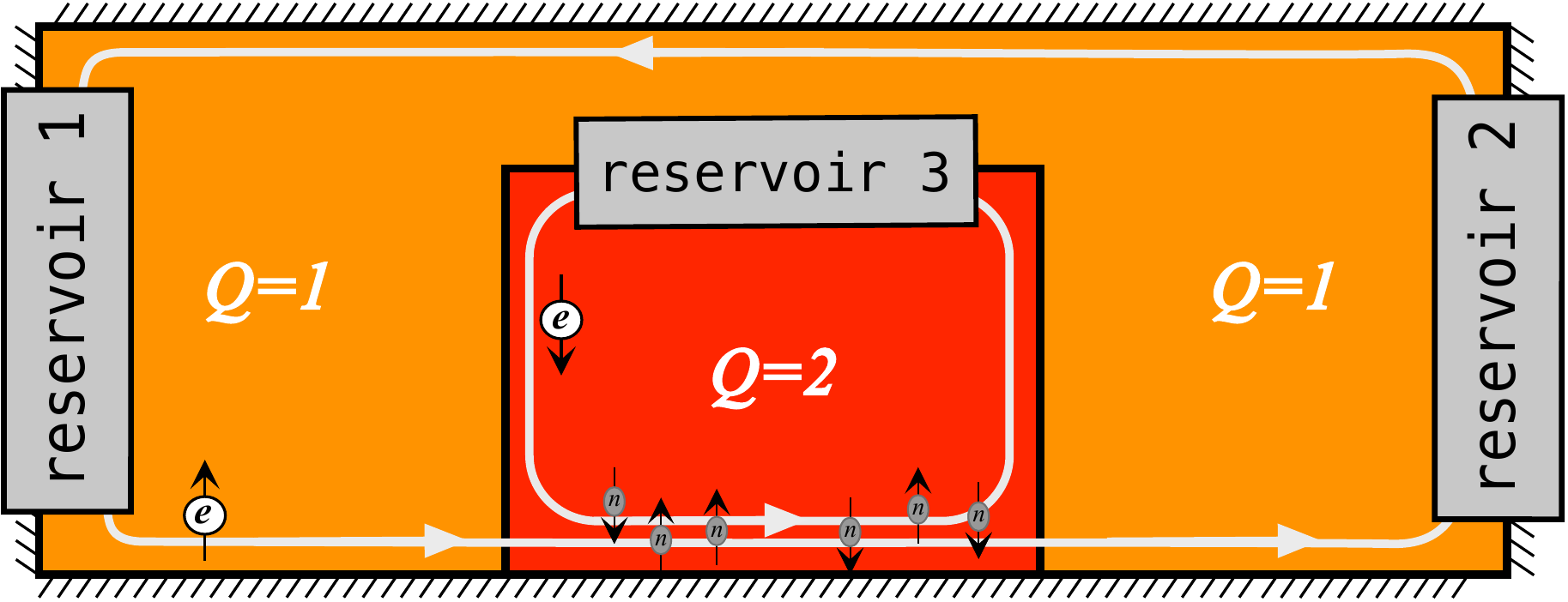}
\caption{Second alternative version of topological information device. In this case, the system is a quantum Hall insulator in a Corbino geometry. The red region has $\mathcal{Q}=2$ with spin-down chiral modes, whereas the orange region has $\mathcal{Q}=1$ with a spin-up chiral mode. One of the chiral modes of the spin-down electrons overlap with the spin-up electrons originating from the outer reservoirs. Two outer terminals are kept at the same chemical potential, whereas a voltage bias is applied between the inner and outer reservoirs. In this configuration, current can only flow from outer to inner terminals only when there is a spin-flip interaction with nuclear spins.}\label{FIG:SuppMat:TID_v3}
\end{figure}

Another alternative version of topological information device can be realized by using a quantum Hall insulator in Corbino geometry as depicted in Fig.~\ref{FIG:SuppMat:TID_v3}. This version is non-local transport setup, i.e. there are three reservoirs connected to the device. Two reservoirs are connected to the outer region and whereas the third reservoir is connected to the inner region shown in red in Fig.~\ref{FIG:SuppMat:TID_v3}. We set the filling of the red region to have $\mathcal{Q} = 2$ with spin-down chiral modes, whereas the filling factor of the rest of the device is set to have $\mathcal{Q} = 1$ with spin-up chiral mode.

This geometry and filling factor configuration allows for one of the chiral modes originating from the third reservoir to extend to the outer edge of the sample. In the region where this chiral mode (with spin-down) meets with the chiral mode circulating at the outer edge of the device (with spin-up). Similar to our original proposal, nuclear spins can mediate a scattering between these two chiral channel via spin-flip mechanism. An applied voltage bias between inner and outer terminals, while keeping the two outer terminals at the same voltage, can only drive a charge current through the system via nuclear spin-flip. As a result, applied voltage bias generates dynamic nuclear spin polarization. Conversely, a finite nuclear spin polarization can induce a charge current in the system by inducing an imbalance between spin up and spin down chiral modes.

\section{Derivation of Maxwell's Demon Induced Current and Generated Power for  Topological Information Device}\label{SuppMat:Derivation}


In this section, we derive the MD induced charge current for topological information device. As a model, we use the second version of topological information device described above. We note that the derivation is more general for the second version whereas the physics and the resulting current voltage relation is the same.

We first write down Boltzmann transport equation for the helical edge electrons in the bottom edge of the device:
\begin{align}
\partial_tf^{\textrm{bot}}_{\uparrow,\downarrow}=
\pm(\Gamma^{\textrm{bot}}_{+-}(\epsilon,
x)-\Gamma^{\textrm{bot}}_{-+}(\epsilon, x)\,)\,\nu(0)^{-1}\mp
v_F\partial_xf^{\textrm{bot}}_{\uparrow,\downarrow},
\end{align}
where $f_\sigma$ is the distribution function for the electrons with spin $\sigma$ and the superscript $bot$ denotes the bottom edge. Here, $\Gamma^{\textrm{bot}}_{-+}$ and $\Gamma^{\textrm{bot}}_{+-}$ are the scattering rates for the electrons in bottom edge, scattering from left to right and from right to left, respectively. We calculate these scattering rates as follows:
\begin{eqnarray}\label{EQN:SpinFlipRate}
\begin{aligned}
\Gamma^{\textrm{bot}}_{-+}(\epsilon,x)= \frac{\gamma_0}{\hbar}\,
N^{\textrm{bot}}_{\downarrow} (x)\,
f^{\textrm{bot}}_{\uparrow}(\epsilon,x)(1-f^{\textrm{bot}}_{\downarrow}(\epsilon,x)),\\
\Gamma^{\textrm{bot}}_{+-}(\epsilon,x)= \frac{\gamma_0}{\hbar}\,
N^{\textrm{bot}}_{\uparrow} (x)\,
f^{\textrm{bot}}_{\downarrow}(\epsilon,x)(1-f^{\textrm{bot}}_{\uparrow}(\epsilon,x)).
\end{aligned}
\end{eqnarray}

We assume that the nuclear polarization $m$ is changing slowly and seek a steady state solution. In this case, distribution functions obey the steady state Boltzmann equation:
\begin{align}
\partial_xf^{\textrm{bot}}_{\uparrow,\downarrow}&=\big(\Gamma^{\textrm{bot}}_{+-}(\epsilon,
x)-\Gamma^{\textrm{bot}}_{-+}(\epsilon,
x)\big)\,\big(v_F\nu(0)\big)^{-1},\nonumber\\
&\equiv\Gamma^{\textrm{bot}}[f^{\textrm{bot}}_{\uparrow},f^{\textrm{bot}}_{\downarrow}],
\end{align}
where $v_F$ is the Fermi velocity and $\nu(0)$ is the density of states. We expand in gradients of the distribution function and obtain a linear dependence in position as:
\begin{equation}
\label{EQN:DistFunc}
f^{\textrm{bot}}_{\uparrow,\downarrow}(x)=f_{L,R}^0+\Gamma^{\textrm{bot}}[f_{L}^0,f_{R}^0]\,(x\pm
L/2),
\end{equation}
where we use the boundary conditions $f^{\textrm{bot}}_{\uparrow}(x = -L/2) = f_{L}^0$ and $f^{\textrm{bot}}_{\downarrow}(x = L/2) = f_{R}^0$. Here, $f^0_L$ and $f^0_R$ are distributions of left and right reservoirs, respectively.

In a similar fashion, we write down the Boltzmann equation for the top edge:
\begin{align}
\partial_tf^{\textrm{top}}_{\uparrow,\downarrow}=
\mp(\Gamma^{\textrm{top}}_{+-}(\epsilon,
x)-\Gamma^{\textrm{top}}_{-+}(\epsilon, x)\,)\,\nu(0)^{-1}\pm
v_F\partial_xf^{\textrm{top}}_{\uparrow,\downarrow},
\end{align}
where the superscript $\textrm{top}$ denotes the top edge and scattering rates for top edge are given as:
\begin{eqnarray}\label{EQN:SpinFlipRatetop}
\begin{aligned}
\Gamma^{\textrm{top}}_{-+}(\epsilon,x)= \frac{\gamma_0}{\hbar}\,
N^{\textrm{top}}_{\uparrow} (x)\,
f^{\textrm{top}}_{\downarrow}(\epsilon,x)(1-f^{\textrm{top}}_{\uparrow}(\epsilon,x)),
\\
\Gamma^{top}_{+-}(\epsilon,x)= \frac{\gamma_0}{\hbar}\,
N^{\textrm{top}}_{\downarrow} (x)\,
f^{\textrm{top}}_{\uparrow}(\epsilon,x)(1-f^{\textrm{top}}_{\downarrow}(\epsilon,x)).
\end{aligned}
\end{eqnarray}

In steady state, the distribution functions at the top edge obey:
\begin{align}
\partial_xf^{\textrm{top}}_{\uparrow,\downarrow}&=\big(\Gamma^{\textrm{top}}_{+-}(\epsilon,
x)-\Gamma^{\textrm{top}}_{-+}(\epsilon,
x)\big)\,\big(v_F\nu(0)\big)^{-1}\nonumber\\
&\equiv\Gamma^{\textrm{top}}[f^{\textrm{top}}_{\uparrow},f^{top}_{\downarrow}].
\end{align}

Before we make use of the same expansion to obtain the distribution functions at the top edge, we revisit the property of topological information device. As only one spin species is present in one of the leads, there is no state within a given lead for the edge states with the opposite spin to occupy. Therefore, the edge state with the opposite spin is extended from bottom edge to the top edge. We express this property as a boundary condition that associates the distributions function of the top and bottom edges:
\begin{align}\label{EQN:2ndBC}
&f^{\textrm{bot}}_{\uparrow}(x = L/2) = f^{\textrm{top}}_{\uparrow}(x=L/2)\\
&f^{\textrm{bot}}_{\downarrow}(x = -L/2) =
f^{\textrm{top}}_{\downarrow}(x=-L/2)
\end{align}

Using Eq.~\eqref{EQN:2ndBC}, we now expand in gradients of the distribution functions at the top edge as follows:
\begin{align}
\label{EQN:DistFunctop}
f^{\textrm{top}}_{\uparrow,\downarrow}(x) &=f^{\textrm{bot}}_{\uparrow,\downarrow}(\pm
L/2)\nonumber\\[3pt]
& +\Gamma^{\textrm{top}}[f^{\textrm{bot}}_{\uparrow}(L/2),f^{\textrm{bot}}_{\downarrow}(-L/2)]\,(x\mp
L/2).
\end{align}

We then use these distribution functions and  obtain the total charge current flowing through topological information device:
\begin{equation}
I_{tot} = \frac{e}{h}\int d\epsilon \big(
f^{\textrm{top}}_{\downarrow} + f^{\textrm{bot}}_{\uparrow} -
f^{\textrm{bot}}_{\downarrow} - f^{\textrm{top}}_{\uparrow} \big)
\end{equation}

Using Eq.~\eqref{EQN:DistFunc} and Eq.~\eqref{EQN:DistFunctop}, we rewrite the equation above as:
\begin{align}\label{EQN:totcur}
I_{tot} &= \frac{e}{h}\int d\epsilon \bigg(\Gamma^{\textrm{top}}[f^{\textrm{bot}}_{\uparrow}(\frac{L}{2}),f^{\textrm{bot}}_{\downarrow}(-\frac{L}{2})] - \Gamma^{\textrm{bot}}[f_{L}^0,f_{R}^0] \bigg) L\nonumber\\ &= e\big(N^{\textrm{top}} \frac{d m^{\textrm{top}}}{dt} + N^{\textrm{bot}} \frac{d m^{\textrm{bot}}}{dt}\big).
\end{align}

Eq.~\eqref{EQN:totcur} shows that the charge current flows through the system only via nuclear spin dynamics. The form of Eq.~\eqref{EQN:totcur} suggests that we can modify the number of nuclear spins for both edges (i.e. $N^{\textrm{bot}(\textrm{top})}$) to get the desired current characteristics. In a device in which there are no nuclear spins present at the top edge, i.e. $N^{\textrm{top}} = 0$, we arrive at the charge current:
\begin{align}
I_{tot} = eN\frac{dm}{dt},
\end{align}
where we dropped the label for the bottom edge for clarity.

We note that in the limit of vanishing junction length for the quantum spin Hall insulator, the number of nuclear spins participating in the spin-flip process changes, however the results remain unchanged. In fact, in this limit, we obtain our first version of topological information device presented at the main text.

In this limit, we obtain the total power generated as:
\begin{align}\label{EQN:pow2}
P_{tot} &= \frac{e^2 V^2}{h}\xi\bigg[\frac{1}{2} - \coth\left(\frac{eV}{2k_B T}\right)m\bigg].
\end{align}
We recall the formula for mean polarization dynamics given by:
\begin{equation}
\label{EQN:mpd}
m(t) = (m_0 - \bar{m})e^{-t/\tau} + \bar{m},
\end{equation}
where $\bar{m} = \frac{1}{2}\tanh\left(eV/2k_B T\right)$ is the target mean polarization and $\tau =(\gamma_0 eV/\hbar)^{-1}\tanh\left(eV/2k_B T\right)$ is the characteristic time scale for mean polarization dynamics. We insert Eq.~\eqref{EQN:mpd} into Eq.~\eqref{EQN:pow2} and obtain the power generated:
\begin{align}\label{EQN:pow3}
P_{tot} &= \frac{e^2 V^2}{h}\xi\bigg[\frac{1}{2} -m_0\coth\left(\frac{eV}{2k_B T}\right)\bigg]e^{-t/\tau}.
\end{align}

\section{Memory Erasure Using Constant Voltage Bias} 
In this section, we show how to erase the memory by applying a constant voltage bias and polarizing nuclear spins. We start from zero mean polarization, $m_0 = 0$, and polarize the nuclear spins up to $\kappa/2$, where $\kappa$ is defined to be the fraction of nuclear spins we choose to polarize:
\begin{equation}
\label{EQN:Fraction}
\frac{\kappa}{2} = \bar{m}(1-e^{-\bar{t}/\tau}).
\end{equation}

Here, $\bar{t} = - \tau \ln(1-\kappa\coth(eV/2k_B T))$ is the time when we stop erasing the memory. We obtain the heat dissipated by taking the integral of power over time:
\begin{align}
W_{C} &= \int_0^{\bar{t}} dt \frac{e^2
V^2}{h}\frac{\xi}{2}e^{-t/\tau}\nonumber\\
& = \frac{eV}{2k_B T}\kappa N k_B T
\end{align}

Note that $0\leq 1- \kappa/2\bar{m}<1$ (see Eq.~\eqref{EQN:Fraction}). Therefore, this conditions sets a lower bound on the applied voltage:
\begin{align}
V &\geq \frac{k_B T}{e} \ln\left(\frac{1 +
\kappa}{1-\kappa}\right).
\end{align}

Taking the lower bound value as the applied voltage bias, we obtain the minimum amount of heat dissipated during the memory erasure phase:
\begin{align}
W_{C} = \frac{\kappa}{2}
\ln\bigg(\frac{1+\kappa}{1-\kappa}\bigg) N k_B T.
\end{align}

\section{Discharging Phase}\label{SuppMat:WorkExtraction}
\subsection{Work extraction under constant voltage bias}
Eq.~\eqref{EQN:pow3} shows that $P_{tot} \leq 0$ for $\frac{1}{2} - m_0\coth\left(\frac{eV}{2k_B T}\right) \leq 0$. We now find the extracted work under constant voltage bias by integrating Eq.~\eqref{EQN:pow3} over time:
\begin{align}
\label{EQN:SuppMat_Work}
W &= \int_0^\infty dt \frac{e^2 V^2}{h}\xi\bigg[\frac{1}{2}
- m_0\coth\left(\frac{eV}{2k_B
T}\right)\bigg]e^{-t/\tau}\nonumber\\
& = \frac{eV}{2k_B T}\bigg( \tanh\left(\frac{eV}{2k_B
T}\right) - 2m_0\bigg) N k_B T
\end{align}

We assume that the nuclear spins are fully polarized initially, ($m_0 = 0.5$). We maximize Eq.~\eqref{EQN:SuppMat_Work} with respect to applied voltage $V$ and find the maximum work can that can be extracted as:
\begin{equation*}
W_{ext} \approx 0.4 Nk_BT \ln(2),
\end{equation*}
for $eV \approx 1.28 k_B T$. This means that,using constant voltage bias, we can extract work equivalent up to $\% 40$ of Landauer limit. 
\subsection{Work extraction using a load resistance}

We now consider the case in which an external load connected to topological information device. We show that work extraction is possible at the Landauer limit, as discussed in the main text, with an appropriate choice of load resistance. The power generated by topological information device for an attached load with conductance $G_L$ is given as:
\begin{align}\label{EQN:Power}
P = G_L V^2 = G_L \bigg[\frac{2 k_B T}{e} \tanh^{-1}\big(\alpha m(t)\big)\bigg]^2,
\end{align}
where we used the induced voltage $V = \frac{2 k_B T}{e} \tanh^{-1}\big(\alpha m(t)\big)$ with $\alpha = \zeta/(G_L/G_0 + \zeta/2)$. The extracted work by topological information device is given as
\begin{align}\label{EQN:Work}
W = \int dt G_L \bigg(\frac{2 k_B T}{e}
\tanh^{-1}\big(\alpha m(t)\big)\bigg)^2.
\end{align}

We now use change of variables to reexpress Eq.~\eqref{EQN:Work} and obtain
\begin{align}\label{EQN:Work2}
W &= \int dm \frac{dt}{dm} G_L \bigg(\frac{2 k_B T}{e} \tanh^{-1}\big(\alpha m(t)\big)\bigg)^2,\nonumber\\ & = \frac{G_L}{G_0} \frac{4k_B T}{2\pi\gamma_0}\frac{1}{\alpha - 2}\int_{\alpha/2}^{0} dx \tanh^{-1}\big(x\big),
\end{align}
where in the last line, we used Eq.~\eqref{EQN:dmdt} and later defined $x = \alpha m(t)$ for convenience. We assumed that all nuclear spins were polarized initially, $m=0.5$, and finally, completely depolarized, $m=0$. This process is reflected in the limits of the integral. Finally, we take the integral in Eq.~\eqref{EQN:Work2} and arrive at the work extracted for a given load conductance $G_L$ as given in Eq.~\eqref{EQN:WorkExtraction} in the main text:
\begin{align}\label{EQN:Work3}
W =   \frac{2Nk_BT}{(2-\alpha)\zeta}\frac{G_L}{G_0}\bigg(\ln\big(1-\frac{\alpha^2}{4}\big)+\alpha\tanh^{-1}\big(\frac{\alpha}{2}\big)\bigg).\end{align}

\begin{thebibliography}{99}
\bibitem{FN:Footnote1}
We note that the current topological information device flows in the opposite direction to the applied voltage bias. Therefore, when we match the total current given in Eq.~\eqref{EQN:dccurrent} with the current on the load, we need to account for the additional minus sign.
\end{thebibliography}
\end{document}